\documentstyle[aps,12pt]{revtex}
\begin{document}
\title{Electronic Structure of the c(2x2)O/Cu(001) System}
\author{Sergey Stolbov$^{(1)}$, Abdelkader Kara$^{(1)}$, 
Talat S. Rahman$^{(1,2)}$}
\address{$^{(1)}$ Department of Physics, Cardwell Hall,
Kansas State University, Manhattan, KS 66506, USA} 
\address{$^{(2)}$ Fritz Haber Institute, Max Planck Gesselschaft,
4-6 Faraday Weg, D-14195 Berlin, FRG}

\maketitle

\begin{abstract}

The locally self-consistent real space multiple scattering technique
has been applied to calculate the electronic structure and chemical
binding for the c(2x2)O/Cu(001) system, as a function of $d_{O-Cu1}$
-- the height of oxygen above the fourfold hollow sites. It is found
that the chemical binding between oxygen and copper has a mixed 
ionic-covalent character for all plausible values of $d_{O-Cu1}$. 
Furthermore, the electron charge transfer from Cu to O depends 
strongly on $d_{O-Cu1}$ and is traced to the variation of the 
long-range electrostatic part of the potential. A competition between the 
hybridization of Cu1-$d_{xz}$ with O-$p_x,p_y$ and Cu1-$d_{x^2-y^2}$ 
with O-$p_z$ states controls modification of the electronic structure 
when oxygen atoms approach the Cu(001) surface. The anisotropy of the 
oxygen valence electron charge density is found to be strongly and 
non-monotonically dependent on $d_{O-Cu1}$. 
\end{abstract}

%%\twocolumn

\section{Introduction}

Oxygen adsorbtion on Cu(001) has been the subject of much discussion 
and debate in the past three decades mainly because of the illusive
nature of the adsorption geometry. The history of the field is nicely
summarized in a recent paper \cite{kittel}. Here we mention some of 
the essential points. Unlike the case of Ni(001) in which a true
$c(2\times2)$ phase is observed as the oxygen coverage approaches
0.5ML \cite{dem70} in which the oxygen atoms occupy fourfold hollow
sites at a distance about 0.8\AA \cite{oed89} above the top Ni layer, 
the stable surface structure for similar oxygen coverage on Cu(001)
has the periodicity $(2\sqrt{2}\times\sqrt{2})R45^{\circ}$
\cite{wuttig} and involves missing rows of Cu atoms according to LEED 
\cite{atr90} and surface $X-$ray diffraction \cite{rob} measurements.
This conclusion was agreed upon after much debate and analysis of
data from several complimentary experimental techniques, some of which
indicated the presence of a $c(2\times2)$ structure on an 
unreconstructed Cu(001) \cite{kono,holland,dobler,sotto}. Other 
experiments have indicated the presence of both of these phases 
\cite{tobin,lederer} and still others have given evidences of 
disordered phases in the system \cite{leib}. A more recent 
scanning tunneling microscopic (STM) study has also affirmed the
presence of $c(2\times2)$ overlayer of adsorbed oxygen on Cu(001) 
at low coverage, but only in nanometer size domains and not  
large well-ordered domains \cite{fo96}.

Currently, based on the above findings, the following picture can 
be built. If the adsorbate coverage is low, oxygen atoms adsorb on 
Cu(001) forming $c(2\times2)$ islands. Increase in the coverage leads
to the ordering of the $c(2\times2)$ domains, forming $c(4\times6)$
pattern. When the coverage approaches a critical value of 34\% of
monolayer, the $(2\sqrt{2}\times\sqrt{2})R45^{\circ}$ structure made
by a missing row type reconstruction is formed.

The competition between the observed O/Cu(001) phases and the 
subsequent reconstruction of the surface with the 
$(2\sqrt{2}\times\sqrt{2})R45^{\circ}$ overlayer shows signs of a
delicate balance in the surface electronic structure of the system 
which tilts in favor of one phase or the other depending on the
coverage. A detailed description of the electronic states and 
chemical bonds formed when oxygen atoms adsorb on the Cu surface 
should thus provide much needed insight into the factors responsible 
for the observed variations in the surface structure.

A possible theoretical explanation follows from effective medium 
theory \cite{nors90} which  predicts a reconstructed stable state 
for the substrate. These researchers find that the O valence levels 
interact very strongly with the Cu $d$-bands such that the states 
can be shifted through the Fermi level. The difference in the total 
energy of the reconstructed and the unreconstructed Cu surface 
arises from dependence of the energetic location of the antibonding 
levels, derived from the O 2$p$ -- Cu 3$d$  hybridization, on 
coordination.  A missing row can force this level up, thereby 
pushing up the hybridized antibonding level which can straggle 
the Fermi level, leading to a reduced occupation and thus stronger 
net bonding. In the case of a Ni substrate the 3$d$ band straggles 
the Fermi level, and so its position is largely pinned by the need 
for charge neutrality.  

The rationale for a reconstructed Cu surface with O as adsorbate 
has also been related to the instability of the 
unreconstructed surface due to the large charge transfer to the 
adsorbate \cite{col,bagus}. Calculations performed for oxygen on 
small clusters of Cu (2--25 atoms \cite{madh} and 5 atoms \cite{bagus}) 
find a strong electronic charge transfer from the metal atoms to 
oxygen in the amount of 1.2e \cite{madh} or 1.5e \cite{bagus}. 
These two factors (the degree of the $p$O -- $d$Cu hybridization and 
amount of the effective charge on O) have thus far been proposed as 
the driving forces for the phase instability of the $c(2\times2)$ 
structure. The {\it ab initio} cluster calculations has also found
the O--Cu bond to be mostly ionic with some participation of the 
$d$-electrons of Cu \cite{bagus}. It should be borne in mind, however 
that models like the effective medium theory are at best 
semi-empirical. Further none of the theoretical methods as far 
employed take into account the contribution of the Madelung potential 
for these two overlayer structures, which may significantly change 
the subband energetic positions and charge distribution. If indeed 
an appreciable charge transfer and details of $p$O -- $d$Cu 
hybridization are responsible for the stability of a superstructure 
on Cu(001), a systematic calculation of these quantities
is warranted. Such a calculation must also provide a 
comprehensive description of the relationship between the 
electronic and geometric structures of the system, including the 
contribution of the Madelung potentials. This paper is an effort
in this direction.

 We have carried out detailed calculations of the electronic 
structure of the O/Cu(001) system using a reliable method and 
by varying the separation between the oxygen overlayer and 
the Cu(001) surface. As we shall see, there continues to be 
a debate also on the height of the oxygen atoms above the 
Cu surface. Our calculational strategy allows us to compare 
the local charge redistribution, modifications of hybridization of 
various electronic states and variations of the local and 
long-range contributions to the potentials, as a function of the 
height of the oxygen overlayer above the surface. It is through 
such systematic studies and, in conjunction with experimental 
observations, that we expect to gain insights into the nature of 
the chemical binding between the O and Cu atoms and establish 
rationale for the formation of a particular overlayer. Since, 
both $c(2\times2)$ and $(2\sqrt{2}\times\sqrt{2})R45^{\circ}$ phases 
have been shown to exist \cite{kittel,fo96}, our focus in this paper
is on the former. At a later stage we will undertake a study of
the missing row reconstructed surface with the large surface 
unit cell to represent the $(2\sqrt{2}\times\sqrt{2})R45^{\circ}$,
which would require a much larger computational effort than
the present one. It should be mentioned that Wiell et al. 
\cite{wiell} have used a realistic full-potential LMTO method 
to examine the hybridization of orbitals of the N overlayer 
on Cu(001). These authors have also touched upon the O/Cu(001) 
system, but no details of the O -- Cu electronic state 
hybridization or the charge transfer have been provided.

The structural studies of the $c(2\times2)$ O/Cu(001) system 
performed with different techniques provide quite different 
results as the interlayer O--Cu spacing ($d_{O-Cu1}$) ranges 
from $0$ to $1.5$\AA. However, a majority of results 
\cite{kittel,dobler,lederer} gives $d_{O-Cu1}$ value in the range 
$0.4-0.8$\AA. The SEXAFS measurements \cite{lederer} suggest
three different configurations. In one of them the O atoms are
located at the fourfold hollow (FFH) sites with $d_{O-Cu1}=0.5$\AA.
Two other configurations represent the distorted FFH geometries
with $d_{O-Cu1}=0.8$\AA. The authors of recent photo-electron
diffraction studies also propose a two-side model involving the O
atoms at FFH positions with $d_{O-Cu1}=0.41$ and $0.7$\AA \cite{kittel}. They
associate these two values with the edge and center positions in
small $c(2\times2)$ domains, respectively. Our calculations are
performed for the exact FFH geometry for a set of $d_{O-Cu1}$
including the values obtained from the different experiments.
We use a computational method based on the theory of multiple
scattering in real space. This allows us to explicitly obtain
the local electronic structure of atoms with different locations.
Our results show a general picture of the electronic state
modification when oxygen atoms approach the Cu(001) surface
and reveal the factors controlling charge transfer and chemical 
bond formation.

\section{Computational Method}

The calculations are based on the density functional theory within
the local density approximation (LDA) \cite{KSh} and multiple
scattering theory. Our approach embodies the local self-consistent
multiple scattering method \cite{Stocks95}, in which a compound is
divided into overlapping clusters called local interaction zones
(LIZs) centered around atoms in different local environments.
For each LIZ,  a system of equations for the $T-$ scattering
matrix in the {\it lattice site -- angular moment} representation
for the {\it muffin-tin} (MT) potential \cite{Stocks80} is solved
self-consistently. The solutions are used to determine the cluster
Green's functions and subsequently the local and total electronic
densities of states and valence charge densities. An application
of this method to materials related to the work here (copper 
oxides) has demonstrated its high efficiency and reliability 
\cite{stolb97,stolb99}.

In our model system, apart from the oxygen overlayer, the top
four Cu layers (Cu1 -- Cu4) were considered to be different from 
the bulk Cu. The fifth and lower Cu layers were supposed to have 
bulk properties. LIZ's were built around nonequivalent atoms 
belonging to the five different layers and contained 71 to 79 
atoms depending on the local configuration. The sizes of the LIZ's 
were taken such that the calculated characteristics of the bulk 
system closely matched those obtained from other reliable methods. 

 At each iteration of the self-consistent process a new potential
is built by solving the Poisson's equation with the charge density
obtained at the previous iteration. To take into account surface
effects we follow the approach developed in Ref. \cite{ujf} where
the MT potential at the $\alpha$th atom belonging to the $i$th
layer is given by
\begin{equation}
V(r_{i\alpha})  =  -\frac{2Z_{i\alpha}}{r_{i\alpha}}+8\pi \left [
\frac{1}{r_{i\alpha}} \int\limits_0^{r_{i\alpha}}\rho_{i\alpha}(x)
x^2dx+\int\limits_{r_{i\alpha}}^{R_{i\alpha}}\rho_{i\alpha}(x)xdx
\right ]
\end{equation}
$$
+ V^{xc}[\rho(r_{i\alpha})]-V^{EC}-V^{xc}(\rho^0)
-4\pi\rho_{i}^{0}+V_{i\alpha}^{Mad}
$$

The nuclear and the MT electronic charge density contributions to
the potential (the first term and the terms in square brackets in
(1), respectively) are of the same form as in the bulk calculations
\cite{janak,schmidt}. The electrostatic $(V^{EC})$ and 
exchange-correlation $(V^{xc}(\rho^0))$ parts of the MT constant 
are taken from the bulk. The last two terms in the above expression 
represent the distant part of the MT potential:
\begin{equation}
V_D=-4\pi\rho_{i}^{0}+V_{i\alpha}^{Mad}
\end{equation}
This part contains explicit information about the surface. The
interstitial charge density $\rho_{i}^{0}$ is supposed to be
layer-dependent reflecting the asymmetry of the system. In
such an approach the space in the vicinity of the surface is
divided by layers belonging to different atomic planes parallel
to the surface. In the present calculations we keep the thickness
of the oxygen layer independent of the interlayer O -- Cu1 spacing
such that the radius of the corresponding Wigner-Seitz sphere is
equal to the O-II ionic radius $R_O=1.24\AA$. The thickness of the
Cu2 and next Cu layers corresponds to the bulk lattice parameter.
The Cu1 layer takes the rest of the space between the O and Cu2
layers. The $\rho_{i}^{0}$ values are obtained by averaging the
interstitial charges. The latter are differences between
Wigner-Seitz $(Q_{i\alpha}^{WS})$ and MT $(q_{i\alpha}^{MT})$ charges,
which are calculated by integration of the electronic charge density
over Wigner-Seitz and MT spheres, respectively. The
$Q_{i\alpha}^{WS}$ values are multiplied by a coefficient to
meet charge neutrality condition for the region disturbed by
the surface. This coefficient approaches unity, with charge 
convergence. The Madelung potential includes monopole 
$M_{i\alpha,j\beta}^{00}$, dipole
$M_{i\alpha,j\beta}^{10}$ and interstitial $V_{ij}[\rho_j]$ terms
\cite{ujf}:

\begin{equation}
V_{i\alpha}^{Mad}=\sum_{j,\beta} \left ( q_{j\beta}M_{i\alpha,
j\beta}^{00}+d_{j\beta}M_{i\alpha,j\beta}^{10}+V_{ij}[\rho_j]
\right ),
\end{equation}
where $d_{j\beta}$ is a dipole moment of the MT charge density and
$$
q_{j\beta}=Z_{j\beta}-q_{j\beta}^{MT}+\frac{4\pi}{3}\rho_jR_{j\beta}^3
$$
The $M_{i\alpha,j\beta}^{00}$ and $M_{i\alpha,j\beta}^{10}$ terms in
(3) are calculated by an Ewald-like plane-wise summation, and explicit
solution of one-dimensional (normal to the surface) Poisson's
equation, using the technique described in Ref. \cite{ujf}.

The exchange and correlation parts of the potential are determined
within LDA using the technique described in Ref. \cite{GunLun}.

To test the applicability of the method to surface problems, we have
calculated the electronic structure of the clean Cu(001) surface and
compared our results with ones obtained by means of TB-LMTO method
\cite{kudr93}, screened KKR Green's function method \cite{wildb} and 
self-consistent localized KKR scheme developed for the surface
calculations \cite{ujf}. As we shall see in the next section, 
our projected densities of electronic states calculated for the
$c(2\times2)$ O/Cu(001) system are in good agreement with those
obtained by full potential LMTO method \cite{wiell}

\section{Results and Discussion}

We have calculated the electronic structure of the O/Cu(001) surface
with the c(2x2) symmetry for selected values of the O-Cu1 interlayer
spacing $d_{O-Cu1}$. As already discussed, there is some discrepancy
in the values of $d_{O-Cu1}$ obtained from different experiments.
However a majority of experiments have found $d_{O-Cu1}$ to lie between 
$0.4\AA$ and $0.8\AA$ ($0.11a-0.22a$, where $a$ is the lattice 
parameter of bulk fcc copper). With this in mind we have varied 
$d_{O-Cu1}$ from $0.15a$ to $0.3a$ in our calculations and examined 
its effect on the surface electronic structure. The local densities 
$N_l^{\alpha}(E)$ of electronic states, their projection 
$N_{lm}^{\alpha}(E)$ on the cubic harmonics, and the valence electron 
density $\rho(r)$ have been calculated for the oxygen overlayer and 
the four top Cu layers, for selected values of $d_{O-Cu1}$.

The densities of the $p$O- and $d$Cu-electronic states of the
system are plotted in Fig. 1 for the longest and shortest interlayer
spacings considered. As seen in Fig. 1, the $N_d^{Cu}(E)$ 
calculated for the Cu3 and Cu4 atoms (third and fourth layer) 
are hardly affected by the choice of $d_{O-Cu1}$ and differ 
slightly from the bulk values whereas the $d$Cu2-states are 
noticeably disturbed. The $d$Cu1 and $p$O state densities 
are found to have a rich structure and for both values of
$d_{O-Cu1}$ the spectra show a clear signature of the 
$p$O -- $d$Cu1 hybridization. Note the position of the peaks
marked  $"a"$ and $"b"$ for $d_{O-Cu1}=0.3a$ and $"a^{\prime} "$ 
and $"b^{\prime} "$ for $d_{O-Cu1}=0.15a$ in Fig. 1. When the oxygen 
overlayer approaches the Cu(001) surface ($d_{O-Cu1} = 0.15a$), 
the splitting (the distance between $a^{\prime}$ -- $b^{\prime}$
peaks in the figure) increases reflecting an enhancement of the
hybridization, which even involves the second copper layer.

To examine the charge transfer in the vicinity of the surface we 
have calculated the Wigner-Seitz charges that allowed us to obtain the
amount of charge deviation ($\Delta Q_i$) from electric neutrality
per $2D-$unit cells belonging to the different layers. The results
are shown in Fig. 2. One can see a strong charge transfer from the
Cu1 layer to the O overlayer coupled with some increase in the 
electron charge in the Cu2 layer. The effect is significantly 
enhanced when the oxygen atoms approach the Cu1 layer. From this 
result one would conclude that the O -- Cu1 chemical binding has an 
essentially ionic character and that the ionicity increases as 
oxygen atoms approach the Cu(001) surface. In Table 1, we present the
amount of charge transfer for the five different heights of the oxygen 
overlayer that we considered. At the shortest O--Cu1 distance, the 
two Cu atoms in the unit cell of the first layer have a deficit of
about one electron each, with 1.5 electron going to the O atom and 
0.25 electron for each of the 2 atoms of the second layer. The amount 
of charge provided to the O atom drops to 1 electron when it's height 
above Cu1 is doubled to 0.3a. The charge deviation $\Delta Q_O$ of the 
O layer and the distant part $V_D$ of the potential in Eq. (2) at the
O site are plotted versus $d_{O-Cu1}$ in the lower panel of Fig. 2.
A clear correlation between $\Delta Q_O$ and $V_D$ seen in the figure
suggests that the long-range electrostatic interaction is a driving
force for the charge transfer in the system.

In the system under consideration, each oxygen atom and its
four nearest Cu1 neighbors form a pyramid. Such an atomic
configuration is expected to cause a strong anisotropy in the
electronic structure. Therefore an analysis of the $N_{lm}^{\alpha}(E)$ 
distributions for O and Cu1 is very useful for understanding the 
nature of the chemical bonds formed when oxygen adsorbs on the Cu 
surface. A simple symmetry consideration shows that only the 
O-$p_x,p_y$ --Cu1-$d_{xz}$ and O-$p_z$ -- Cu1-$d_{x^2-y^2}$ 
hybridization can be significant in the O -- Cu1 subsystem. 
Therefore our focus is mainly on these electronic states. In 
Fig. 3, the densities of the O-$p_z$ and Cu1-$d_{x^2-y^2}$ states 
are plotted for the values of $d_{O-Cu1}$ as labeled. Both 
projected densities are considerably modified with the decrease 
in $d_{O-Cu1}$. The band broadens and distinct new peaks
appear as $d_{O-Cu1}$ takes the values $0.2a$ and $0.15a$. An
energetic alignment of the O-$p_z$ and Cu1-$d_{x^2-y^2}$ peaks
in the $a$, $b$ and $c$ regions is a signature of hybridization
of these states. An increase in the peak amplitudes and an extra
splitting with the decrease in $d_{O-Cu1}$ reflect a strong
enhancement of the hybridization as the oxygen atoms approach
the surface. In contrast, the splitting of the O-$p_x,p_y$
and Cu1-$d_{xz}$ states is almost independent of the $d_{O-Cu1}$
value (see Fig. 4). This means that the hybridization of these
states is not changed noticeably, when $d_{O-Cu1}$ is varied within
the range that we considered. Such a difference in the response 
of the electronic states to variation of the O-Cu1 interlayer 
spacing can be explained on the basis of a simple geometric 
analysis. When the oxygen atoms approach the Cu1 layer, the 
O -- Cu1 bond length decreases and the symmetry of the O-$p_z$ 
and Cu1-$d_{x^2-y^2}$ cubic harmonics is such that their overlap 
increases enhancing the hybridization. On the other hand, the
symmetry of the O-$p_x,p_y$ and Cu1-$d_{xz}$ states is such that 
the overlap of their cubic harmonics is diminished with decrease 
in $d_{O-Cu1}$ that compensates the effect of the O -- Cu1 bond 
length reduction.

Thus, according to these results, a noticeable covalence $p$O --
$d$Cu1 binding takes place in the system along with the strong
ionic binding. If the O -- Cu1 interlayer spacing is long enough
($d_{O-Cu1}=0.3a$), the covalence component of chemical binding
is mostly determined by the O-$p_x,p_y$ -- Cu1-$d_{xz}$
hybridization. When $d_{O-Cu1}$ is smaller, the O-$p_z$ --
Cu1-$d_{x^2-y^2}$ contribution to the covalence bond increases.
This competition mostly controls modification of the electronic
states when the oxygen atoms approach the surface.

The densities of O-$p_x,p_y$ and O-$p_z$ electronic states have 
also been calculated for O/Cu(001) by means of the full-potential 
LMTO method \cite{wiell}, which is considered to be one of the best 
approaches for electronic structure calculations. These authors 
considered an oxygen overlayer with a $c(2\times2)$ structure with 
the oxygen atom occupying an FFH and a quasi-FFH sites.
In Fig. 5 we have compared our results obtained for $d_{O-Cu1}=$ 
0.15a (0.514\AA) with the results of Ref. \cite{wiell} at the
FFH site at a height of 0.5\AA, for O-$p_z$ and O-$p_{x,y}$. 
Both the splitting and peak amplitudes are in very good 
agreement attesting to the reliability of the method used here.
We note from Fig. 5 that a shift the height
between 0.15a and 0.2a produces a shift of about 1eV of the spectrum,
again in good agreement with the results in Ref. \cite{wiell}.

We have calculated the radial dependence of the valence
electron charge density around the O and Cu1 -- Cu4 atoms along
some symmetrical directions. The plot of the oxygen charge density
in Fig. 6 displays it to be essentially anisotropic. We have
also found that the anisotropy increases when $d_{O-Cu1}$ is 
reduced from $0.3a$ to $0.2a$ whereas further $d_{O-Cu1}$
reduction leads to a decrease in the anisotropy. This effect 
is illustrated in Fig. 7, in which the radial dependence of the
difference between the densities along the (001) and (100)
directions is plotted for systems with different values of
$d_{O-Cu1}$. Such a non-monotonic behavior can be explained 
by the hybridization competition mentioned above. The change 
in the anisotropy of the oxygen charge density occurs because 
different amounts of charge flow to the O-$p_x$ and O-$p_z$ 
states, as $d_{O-Cu1}$ decreases. The stronger covalence binding,
the larger an increase in the electron-electron repulsion 
caused by an extra electronic charge in the bond region. 
Therefore the charge tends to come to the states less involved 
in hybridization, which are the O-$p_z$ states when 
$d_{O-Cu1}$ is large. This leads to an increase in the 
anisotropy. The tendency is changed at short interlayer 
distances, when the hybridization of the O-$p_z$ 
and Cu1-$d_{x^2-y^2}$ is enhanced significantly that can 
make preferable an occupation of the O-$ p_x,p_y$ states.

We now turn to the implication of the results presented here 
to a simple model proposed for the adsorption of oxygen on Cu(001). 
To describe this phenomenon, the authors of Ref. \cite{lederer}
use a model isotropic potential based on the local part of the 
Madelung potential. At the same time they assume that not only 
the O and Cu1 layers are involved in charge transfer, but also 
deeper ones. Our results confirm the latter assumption (see 
Fig. 2) and imply that an accurate model potential should be 
based on the long range Coulomb interaction rather than on the 
local one. Furthermore, using the experimental data obtained 
for the $c(2\times2)$ and $(2\sqrt{2}\times\sqrt{2})R45^{\circ}$ 
phases the authors of Ref. \cite{lederer} build the isotropic 
potential, which appears to have step-like dependence on the 
O -- Cu1 bond length. To obtain this, these authors attribute 
a step-like behavior to the effective atomic charges. They 
assume that such a behavior can be induced by an increase in 
covalence, as the O -- Cu1 bond length is reduced. Our results 
do not support these assumptions. We have found that the 
charges and covalence of the bonds are smooth functions of 
the interatomic distance. The $c(2\times2)$ and 
$(2\sqrt{2}\times\sqrt{2})R45^{\circ}$ phases are different 
in the symmetry of the oxygen local surrounding rather than 
in the interatomic distances. This suggests that the phase 
transition can be properly described only by means of an 
anisotropic potential. Moreover, a noticeable covalence of 
the O -- Cu1 binding (even for relatively long bond lengths) 
and its anisotropy obtained in our study allows us to propose 
that an accurate modelling of the $c(2\times2)$ phase itself 
requires an anisotropic potential.

\section{Conclusions}
The electronic structure has been calculated for the O/Cu(001)
system in the $c(2\times2)$ with five different O -- Cu1 interlayer spacings. 
We find that the oxygen and copper atoms form a mixed
ionic-covalence chemical binding in the O/Cu(001) system for
the whole range of the considered O -- Cu1 interlayer spacings.
The oxygen atom gains an extra charge varying between 1 and 
1.5 electron when its height is varied between 0.3a and 0.15a.
This charge comes solely from the atoms at the top layer of
the substrate with a small enhancement of the second layer
 electronic charge.
 We also find that this charge transfer is controlled by variation
of the long-range electrostatic interaction. The detailed
analysis of the properties of the local electronic structure
reveals that the  
electronic structure of the O/Cu(001) system is governed
by a competition between the hybridization of Cu1-$d_{xz}$
-- O-$p_x,p_y$ and Cu1-$d_{x^2-y^2}$ -- O-$p_z$ states, which
depends on  O -- Cu1 spacing. The anisotropy of the oxygen
valence electron charge density is found to be strongly and
non-monotonically dependent on the interlayer spacing. 

{\bf Acknowledgements.}

This work was supported by the National Science Foundation
(Grant No CHE9812397). The calculations were performed on the
Origin 2000 supercomputer at the National Center for
Supercomputing Applications, University of Illinois at
Urbana-Campaign under Grant No DMR010001N. The work of TSR
was also facilitated by the award of an Alexander von
Humboldt Forschungspreis.

\newpage

\begin{figure}
\caption
{The density of $p$-electronic states of the oxygen overlayer and
$d$-electronic states of the four top Cu layers for O/Cu(001)
for O--Cu1 interlayer spacings of 0.3a (dashed line) and 0.15a 
(solid line).}
\end{figure}
\begin{figure}
\caption
{Upper panel: the layer charges per 2D-unit cell. The layer \# 0
corresponds to the oxygen overlayer. Lower panel: charge deviation
for the oxygen overlayer and the distant part of the MT potential
plotted versus the O--Cu1 interlayer spacing.}
\end{figure}
\begin{figure}
\caption
{The densities of the O-$p_z$ (dashed line) and Cu1-$d_{x^2-y^2}$
(solid line) electronic states for $c(2\times2)$ O/Cu(001) with 
different O--Cu1 interlayer spacings.}
\end{figure}
\begin{figure}
\caption
{The densities of the O-$p_{x,y}$ (dashed line) and Cu1-$d_{xz}$
(solid line) electronic states for $c(2\times2)$ O/Cu(001) with 
different O--Cu1 interlayer spacings.}
\end{figure}
\begin{figure}
\caption
{The densities of the O-$p_{x,y}$ and O-$p_z$ electronic states 
calculated in the present work for two O heights (solid line 0.541 \AA) 
and (long-dash dotted line 0.722 \AA) compared with those 
in Ref. 18 (dashed line) for $c(2\times2)$ O/Cu(001) at 0.5 \AA.
A convalution with a lorentzian with 0.7 eV full width at its
half maximum is used.} 
\end{figure}
\begin{figure}
\caption
{The oxygen valence electron charge densities plotted along the
(001) (dashed line) and (100) (solid line) directions for
O/Cu(001) with two different O--Cu1 interlayer spacings.}
\end{figure}
\begin{figure}
\caption
{The radial dependence of the difference between the oxygen
valence electron charge densities calculated along the
(001) and (100) directions for O/Cu(001) with three different
O--Cu1 interlayer spacings.}
\end{figure}

\begin{table}
\caption{Valence electronic charge variation $\Delta Q(e)$ for 
different O-Cu1 interlayer spacing. The percentage change is  
given in parentheses}
\label{tab:val}
\begin{tabular}{cccccc}
\multicolumn{1}{l}{Layer}&
\multicolumn{1}{c}{0.15a}&
\multicolumn{1}{c}{0.17a}&
\multicolumn{1}{c}{0.20a}&
\multicolumn{1}{c}{0.25a}&
\multicolumn{1}{c}{0.30a} \\
\hline
O&1.493 ($37.3\%$)& 1.429 ($35.7\%$)& 1.329 ($33.2\%$)&1.144 ($28.6\%$)&0.983 ($24.6\%$)\\
Cu1&-2.108 ($-9.6\%$)& -1.986 ($-9.0\%$)& -1.800 ($-8.2\%$)&-1.484 ($-6.8\%$)&-1.211 ($-5.5\%$)\\
Cu2&0.575 ($2.6\%$)& 0.519 ($2.4\%$)& 0.430 ($2.0\%$)&0.313 ($1.4\%$)&0.207 ($0.9\%$)\\
Cu3&0.040 ($0.2\%$)& 0.039 ($0.2\%$)& 0.041 ($0.2\%$)&0.027 ($0.1\%$)&0.021 ($0.1\%$)\\
Cu4&0.000 ($0.0\%$)& 0.000 ($0.0\%$)&  0.000 ($0.0\%$)&0.000 ($0.0\%$)&0.000 ($0.0\%$) \\
\end{tabular}
\end{table}


\begin{references}
%%\begin{thebibliography}{99}
\bibitem{kittel} M. Kittel, M. Polcik, R. Terborg, J.-T. Hoeft,
P. Baumg\"artel, A. M. Bradshaw, R. L. Toomes, J.-H. Kang,
D. P. Woodruff, M. Pascal,  C. L. A. Lamont, and E. Rotenberg,
{\it Surf. Sci.} {\bf 470}, 311 (2001).
\bibitem{dem70} J. E. Demuth, D. W. Jepsen, P. M. Marcus,
{\it Phys. Rev. Lett.} {\bf 31}, 540 (1970).
\bibitem{oed89} W. Oed, H. Lindner, U. Starke, K. Heinz,
K. M\"uller, {\it Surf. Sci.} {\bf 224}, 179 (1989).
\bibitem{wuttig} M. Wuttig, R. Franchy, and H. Ibach,
{\it Surf. Sci.} {\bf 213}, 103 (1989).
\bibitem{atr90} A. Atrie, U. Bardi, G. Casalone, G. Rovida,
E. Zanazzi, {\it Vacuum} {\bf 41}, 333 (1990).
\bibitem{rob} I. K. Robinson, E. Vlieg, and S. Ferrer,
{\it Phys. Rev.} B {\bf 42}, 6954 (1990).
\bibitem{kono} S. Kono, S. M. Goldberg, N. F. T. Hall, and
C. S. Fadley, {\it Phys. Rev. Lett.} {\bf 41}, 1831 (1978).
\bibitem{holland} S. P. Holland, B. J. Garrison, and N. Winograd,
{\it Phys. Rev. Lett.} {\bf 43}, 220 (1979).
\bibitem{dobler} U. D\"obler, K. Baberschke, J. St\"ohr, and
D. A. Outka, {\it Phys. Rev.} B {\bf 31}, 2532 (1985).
\bibitem{sotto} M. Sotto, {\it Surf. Sci.} {\bf 260}, 235 (1992).
\bibitem{tobin} J. G. Tobin, L. E. Klebanoff, D. H. Rosenblatt,
R. F. Davis, E. Umbach, A. G. Baca, D. A. Shirley, Y. Huang,
W. M. Kang, and S. Y. Tong, {\it Phys. Rev.} B {\bf 26}, 7076
(1982).
\bibitem{lederer} T. Lederer, D. Arvanitis, G. Comelli,
L. Tr\"oger, and K. Baberschke, {\it Phys. Rev.} B {\bf 48},
15390 (1993).
\bibitem{leib} F. M. Leibsle, {\it Surf. Sci.} {\bf 337}, 51 (1995).
\bibitem{fo96} T.  Fujita, Y. Okawa, Y. Matsumoto, and K. I. Tanaka,
{\it Phys. Rev.} B {\bf 54}, 2167 (1996).
\bibitem{nors90} K. W. Jacobsen and J. K. Norskov
{\it Phys. Rev. Lett.} {\bf 65}, 1788 (1990).
\bibitem{col} E. A. Colbourn, J. E. Inglesfield
{\it Phys. Rev. Lett.} {\bf 66}, 2006 (1991).
\bibitem{bagus} P. S. Bagus and F. Illas {\it Phys. Rev.} B
{\bf 42}, 10852 (1990).
\bibitem{madh} P. V. Madhavan and M. D. Newton {\it Chem. Phys.}
{\bf 86}, 4030 (1987).
\bibitem{wiell} T. Wiell, J. E. Klepeis, P. Bennich,
O. Bj\"orneholm, N. Wassdahl, and A. Nilsson {\it Phys. Rev.} B
{\bf 58}, 1655 (1998).
\bibitem{KSh} W. Kohn, L. J. Sham, {\it Phys. Rev.} {\bf 140},
1133 (1965).
\bibitem{Stocks95} Y. Wang, G. M. Stocks, W. A. Shelton,
D. M. C. Nicholson, Z. Szotek, W. M. Temmerman, {\it Phys. Rev. Lett.}
{\bf 75}, 2867 (1995).
\bibitem{Stocks80} J. S. Faulkner and G. M. Stocks, {\it Phys. Rev.} B
{\bf 21}, 3222 (1980).
\bibitem{stolb97} S.V. Stolbov, {\it J. Phys.: Condensed Matter}
{\bf 9} 4691 (1997)
\bibitem{stolb99} S. Stolbov, M. Mironova, K. Salama, {\it Supercond.
Sci. Technol.} {\bf 12} 1071 (1999).
\bibitem{ujf} L.Szunyogh, B.Ujfalussy, P.Weinberger, J.Kollar,
{\it Phys.Rev.} B {\bf 49}, 2721 (1994).
\bibitem{janak} J. F. Janak, {\it Phys. Rev.} B {\bf 9} 3985 (1974).
\bibitem{schmidt} P. C. Schmidt, {\it Phys. Rev.} B {\bf 31} 5015
(1985).
\bibitem{GunLun} O. Gunnarsson, B. I. Lundqvist, {\it Phys. Rev.} B
{\bf 13}, 4274 (1976).
\bibitem{kudr93} J. Kudrnovsky, I. Turek, V. Drchal, P. Weinberger,
S. K. Bose, and A. Pasturel, {\it Phys.Rev.} B {\bf 47}, 16525 (1993).
\bibitem{wildb} K. Wildberger, R. Zeller, and P. H. Dederichs, 
{\it Phys.Rev.} B {\bf 55}, 10074 (1997).
%%\end{thebibliography}
\end{references}
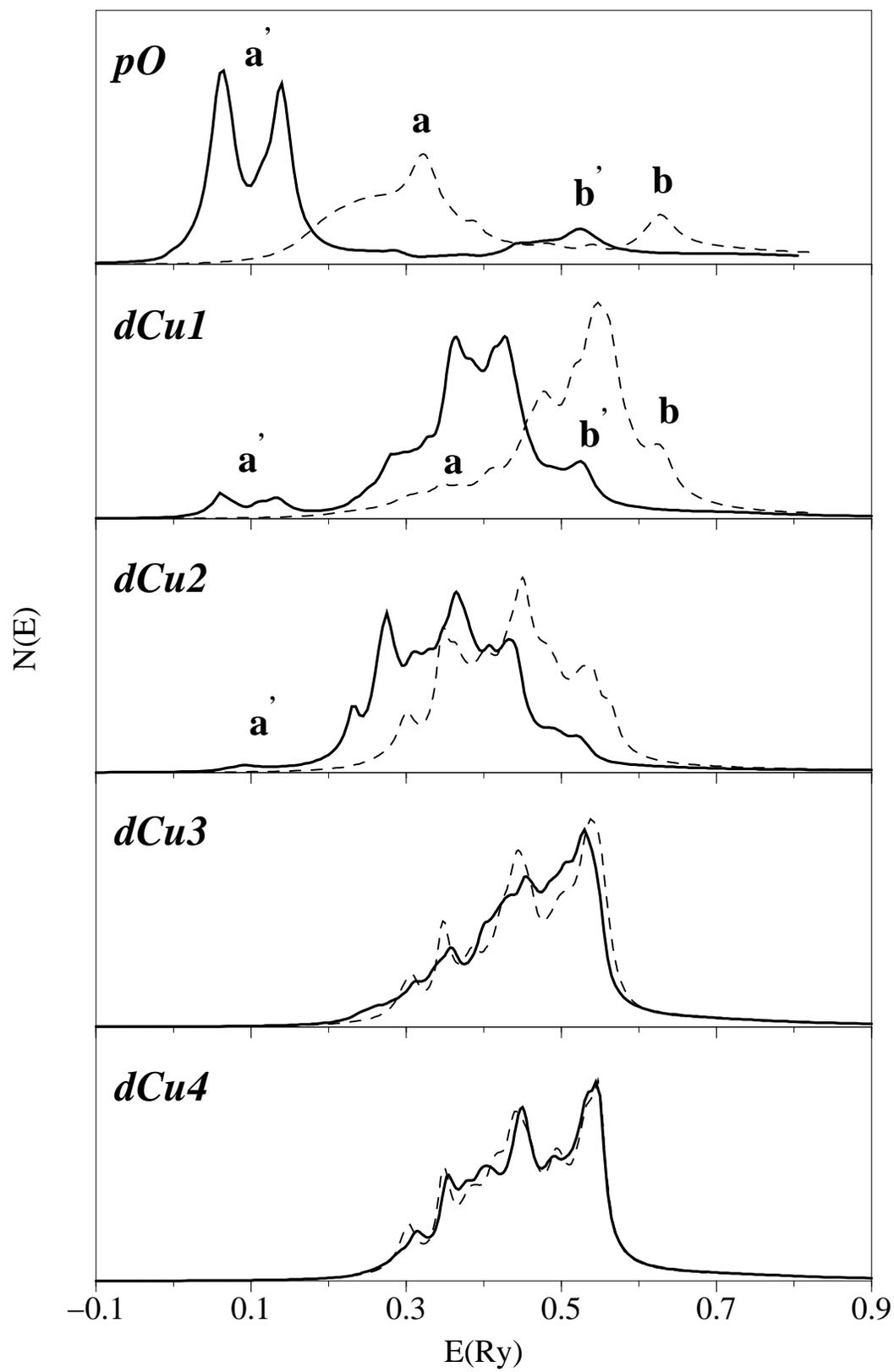
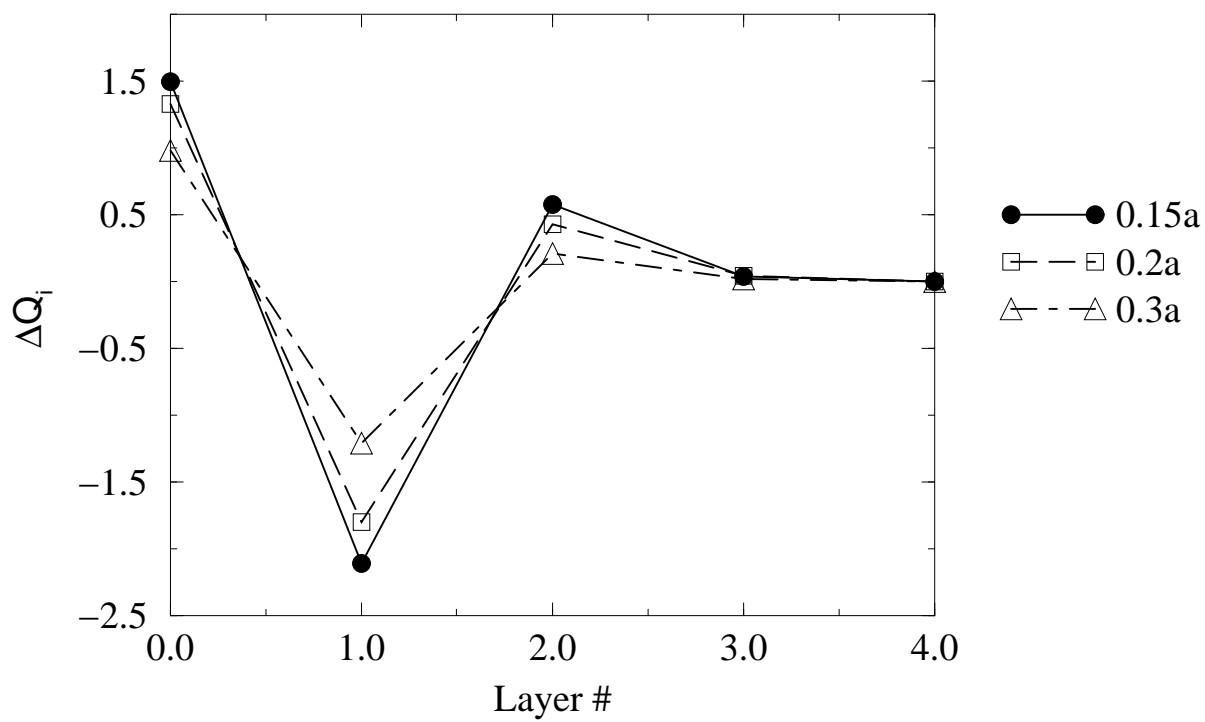
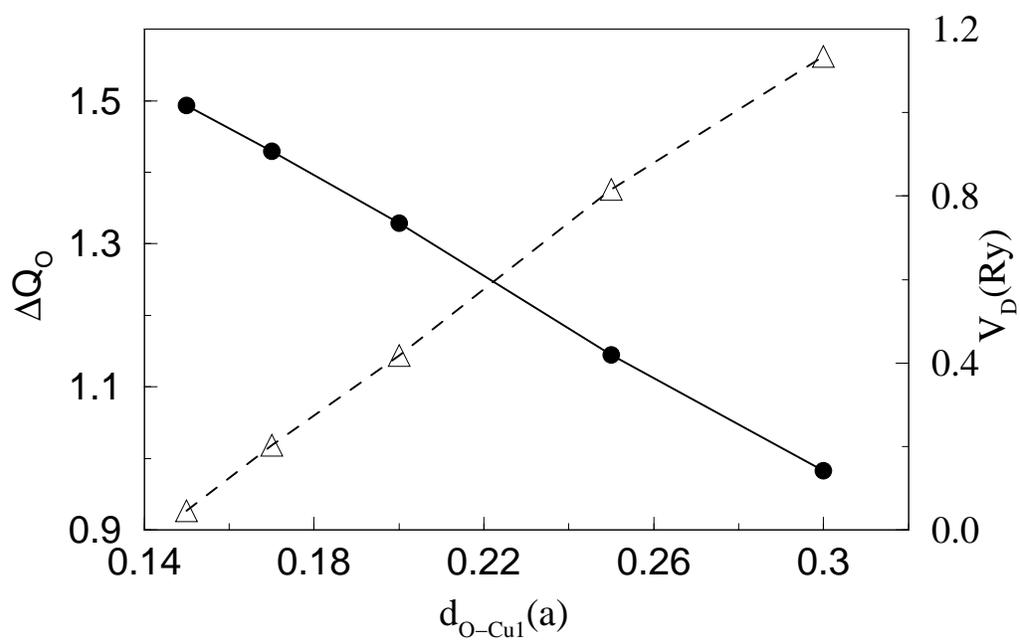
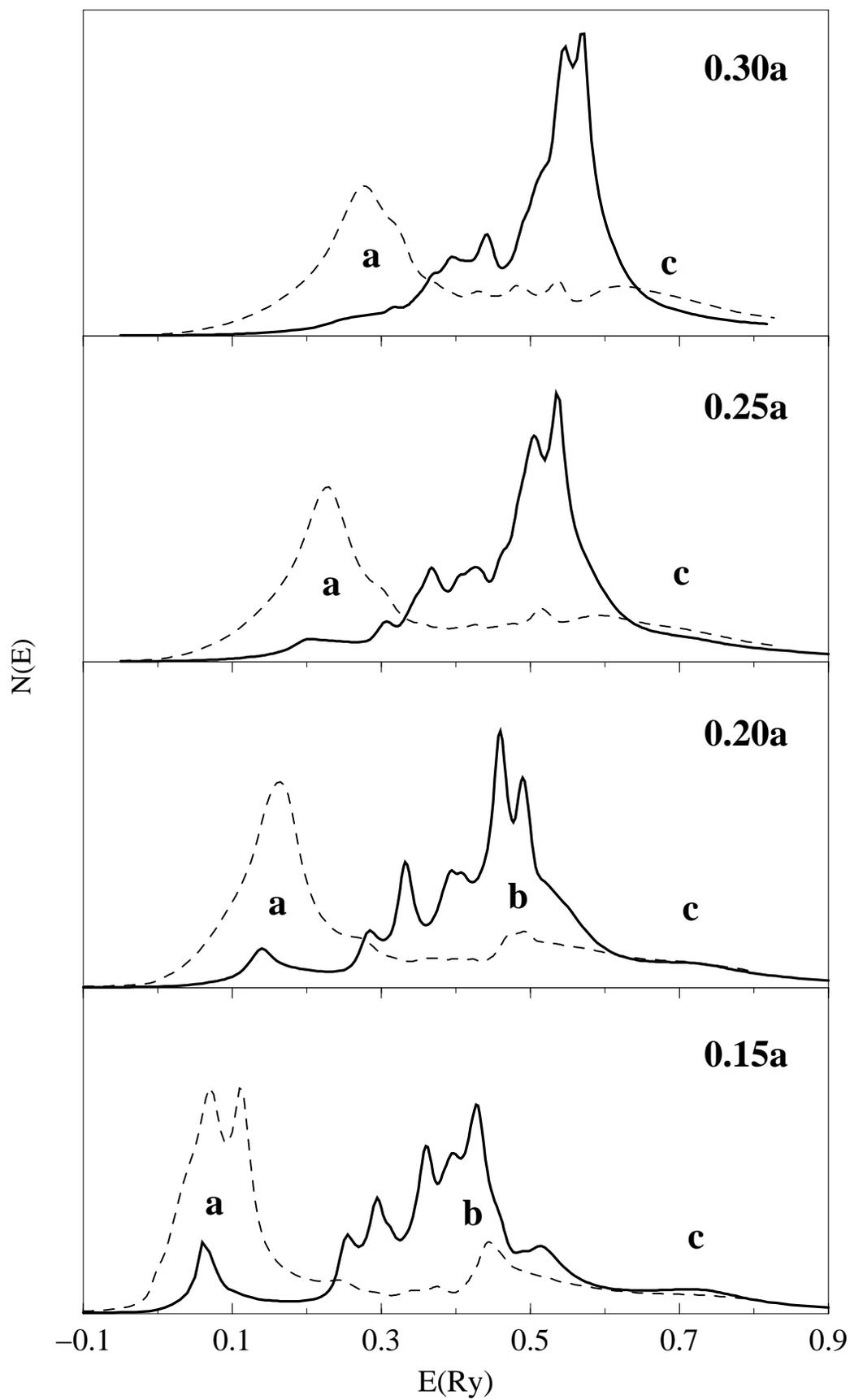
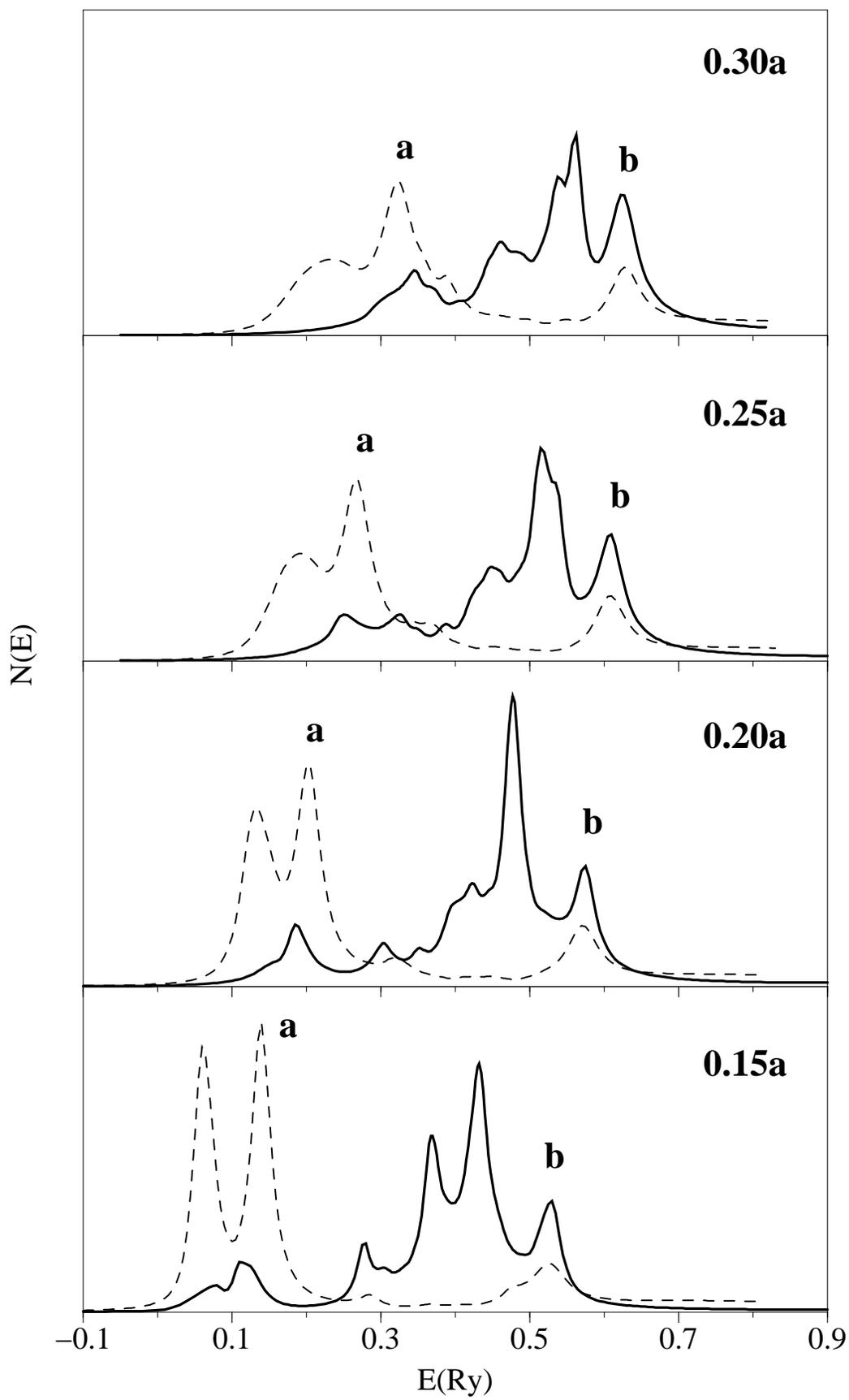
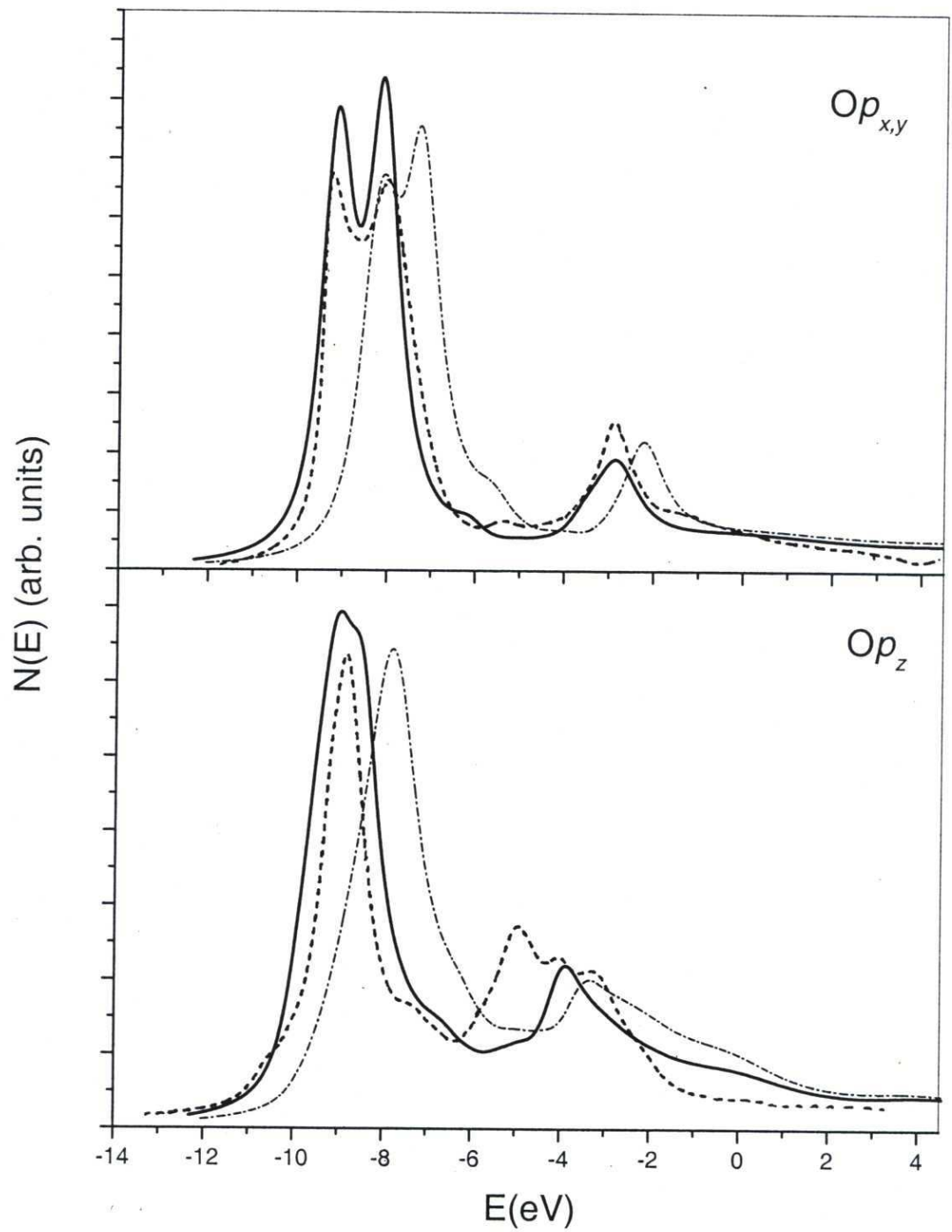
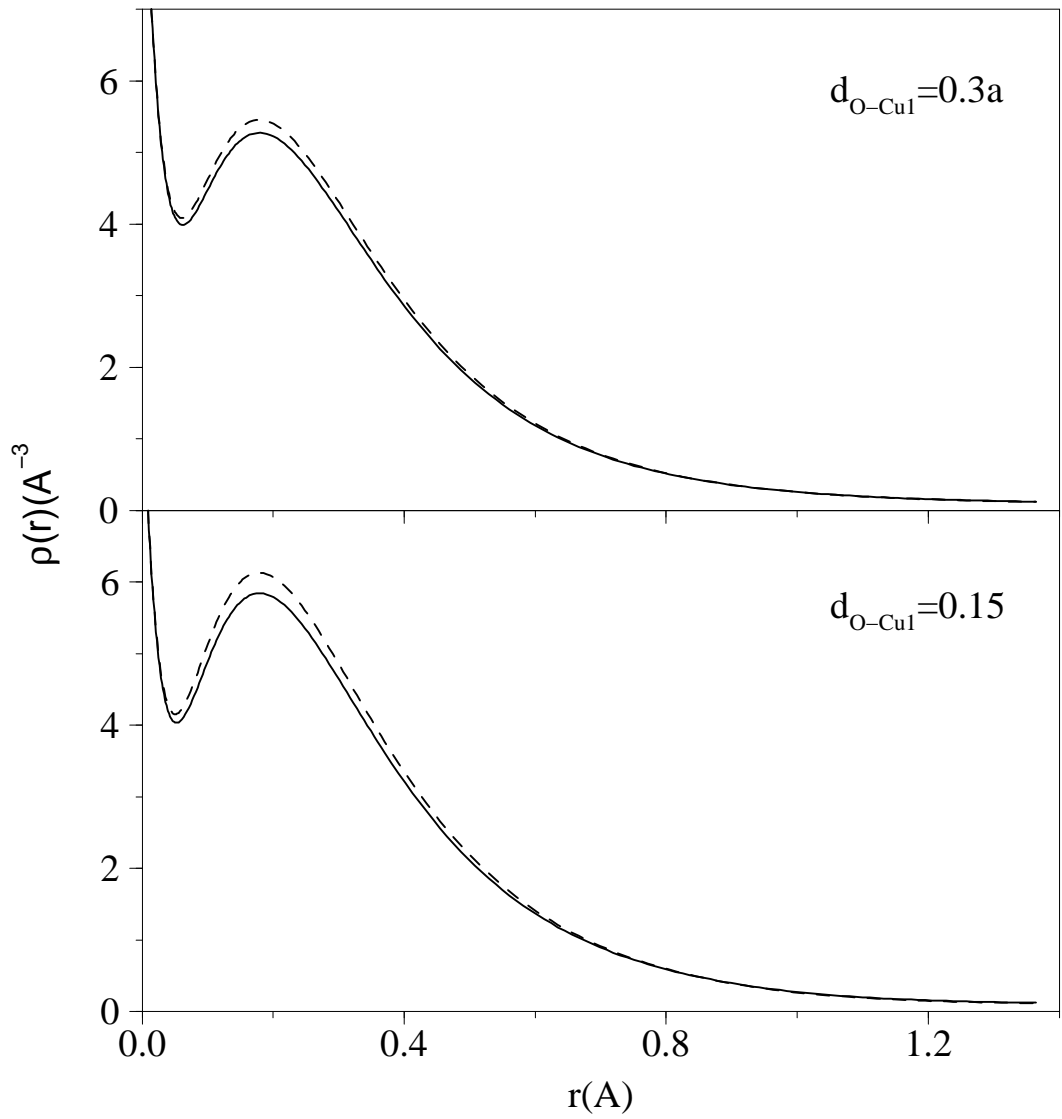
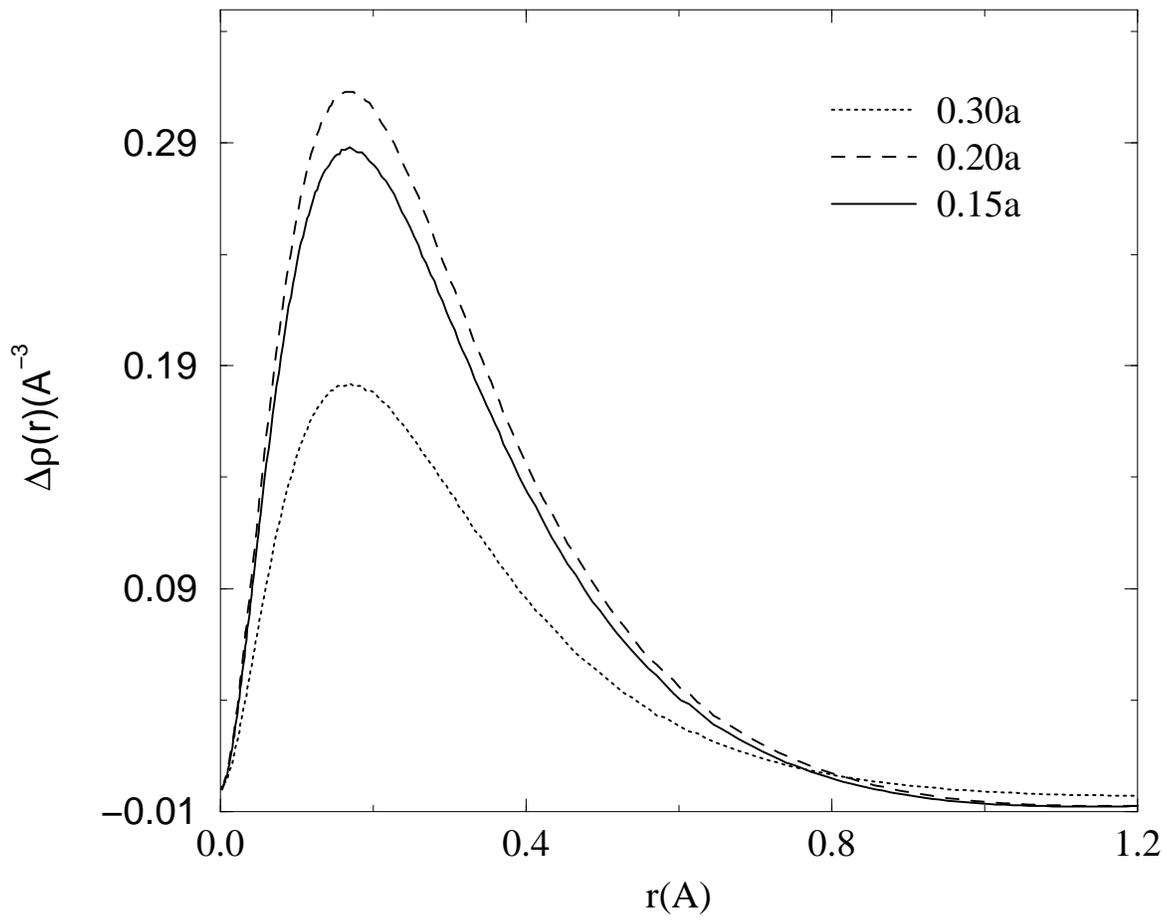
\end{document}